\documentclass[12pt]{article}
\usepackage{fullpage,epsfig,graphics,amsbsy,amssymb,cancel,slashed,mathrsfs}
\usepackage{psfrag,hyperref}
\usepackage{graphicx,color}
\usepackage{wrapfig}
\usepackage{subfigure}
\usepackage{enumerate}

\newcommand{\be}{\begin{eqnarray}}
\newcommand{\ee}{\end{eqnarray}}
\usepackage{amsmath}
\numberwithin{equation}{section}
\usepackage{caption}
\usepackage[nosort]{cite}
 \usepackage[parsep]{collref}

\def\OO{{\mathcal O}}
\def\Z{{\mathbb Z}}
\def\Q{{\mathbb Q}}
\def\CP{\mathbb P}
\def\R{{\mathbb R}}
\def\F{{\mathbb F}}

\newcommand{\bea}{\begin{eqnarray}}
\newcommand{\eea}{\end{eqnarray}}  
\newcommand{\nn}{\nonumber}

 \renewcommand{\Re}{{\tt Re}}

\def\eps{\epsilon}

\begin{document}

\thispagestyle{empty}
\begin{flushright} \small
UUITP-33/19
 \end{flushright}
\smallskip
\begin{center} \LARGE
{\bf Notes on anomalies, elliptic curves}\\
{\bf and the BS-D conjecture}
 \\[12mm] \normalsize
{\bf  Yongchao L\"u and Joseph A. Minahan} \\[8mm] 
 {\small\it
  Department of Physics and Astronomy,
     Uppsala university,\\
     Box 516,
     SE-751 20 Uppsala,
     Sweden\\
   }
  
  \medskip 
   \texttt{  yongchao.lu \& joseph.minahan@physics.uu.se}

\end{center}
\vspace{7mm}
\begin{abstract}
 \noindent   \noindent  We consider anomaly cancellation for $SU(N)\times SU(2)\times U(1)$ gauge theories where the left-handed chiral multiplets are in higher $SU(2)$ representations.  In particular,  if the left-handed quarks and leptons transform under the triplet representation of $SU(2)$ and if the $U(1)$ gauge group is compact then up to an overall scaling there is only one possible nontrivial assignment for the hypercharges if $N=3$, and two if $N=9$.  Otherwise there are infinitely many.  We use the Mordell-Weil theorem, Mazur's theorem and the Cremona elliptic curve database which uses Kolyvagin's theorem on the Birch Swinnerton-Dyer conjecture to prove these statements.
\end{abstract}

\bigskip\bigskip\bigskip\bigskip\bigskip

\centerline{\it In memory of P.~G.~O. Freund}
\eject
\normalsize

\tableofcontents

\section{Introduction}

It has been long known that anomalies  restrict the matter content in particle physics \cite{Bouchiat:1972iq,Gross:1972pv,Georgi:1972bb,AlvarezGaume:1983ig}.  In particular, the ratios of the $U(1)$ hypercharges of the standard model are almost fixed by the cancellation of anomalies \cite{Geng:1988pr}.  The types of anomalies encountered are  $U(1)^3$ anomalies \cite{Adler:1969gk,Bell:1969ts}, mixed anomalies of the $U(1)$ with the other gauge groups \cite{Gross:1972pv}, $SU(3)^3$ anomalies \cite{Georgi:1972bb}, and mixed $U(1)$ gravitational anomalies \cite{Delbourgo:1972xb,Eguchi:1976db,AlvarezGaume:1983ig}.  Assuming that there are five overall hypercharge assignments for the left- and right-handed quarks and leptons, the anomaly cancellation conditions lead to four homogeneous equations for the hypercharges.  Up to an overall scaling there is a zero-dimensional space of solutions where one finds two independent solutions, the hypercharges of the standard model, and a second, almost trivial solution \cite{Geng:1988pr,Minahan:1989vd,Weinberg:1996kr}\nocollect{Weinberg:1996kr}.     Both solutions are rational,  consistent with having the Abelian gauge group be the compact $U(1)$ and not $\R$.  Since one of the anomaly cancellation equations is cubic in the hypercharges, it was not {\it a priori} guaranteed that any solutions would be rational.

Recently Lohitsiri and Tong turned this question around and asked what would happen if instead of requiring the cancellation of the mixed gravitational anomalies, one started with a compact $U(1)$ {\it ab initio}, forcing all charge ratios to be rational \cite{Lohitsiri:2019fuu}.  With one less anomaly equation there is a one-dimensional space of solutions over $\R$.  However, over the rationals, $\Q$, the solutions are discrete.  Since the  $U(1)^3$ anomaly equation is cubic in the hypercharges and the two mixed $U(1)$ gauged anomaly equations are linear, the resulting space is an elliptic curve in $\CP^2$ for complex hypercharges.  Using the linear equations to remove two of the hypercharges and after a convenient linear transformation to a new set of variables the anomaly cancellation equations reduce to the elliptic curve \cite{Lohitsiri:2019fuu}
\be\label{cubic}
X^3+Y^3+Z^3=0\,.
\ee
Up to an overall scaling this equation only has the solutions $(X,Y,Z)=(1,-1,0)$ and its permutations in $\Q$.  Two of these solutions map to the standard model charges and one to the almost trivial solution.  Hence requiring a compact $U(1)$ leads to the same solutions as requiring the cancellation of the  mixed gravitational anomalies.  
This conclusion is  independent of the number of colors $N$, in the sense that choosing either compactness of the $U(1)$ or cancellation of the mixed gravitational anomalies yields the same result: an interesting standard model-like solution and an almost trivial one.

In this note we consider a generalization of the Standard Model generations in order to include more hypercharges.  We do this by assuming that the left-handed quarks and leptons transform in a higher representation $\bf q$ of the $SU(2)$ gauge group.  This requires adding more right-handed quarks to cancel the $SU(N)^3$ anomalies, leading to more hypercharges but the same number of anomaly cancellation equations.  The space of solutions for the hypercharges is then $q-2$ dimensional in $\R$.  If we now assume that the $U(1)$ is compact then the problem is reduced to finding rational solutions on this space.  

Here we focus on  $q=3$ where the left-handed quarks and leptons transform under the triplet of $SU(2)$. Of course this model is not physically realistic, but it turns out to have interesting mathematics associated with it.   In this case the anomaly equations, including the mixed gravitational anomalies, lead to an $N$-dependent elliptic curve in $\CP^2$.  Over the last decades there has been significant progress in understanding the rational solutions for non-singular elliptic curves, including whether or not a curve has a finite number of rational solutions.  Using some of these results we argue that  for $N=3$   there is one nontrivial solution and one almost trivial solution, while for $N=9$ there are two inequivalent nontrivial solutions and one almost trivial solution,  up to relabelings.  For all other values of $N$  there are infinitely many inequivalent rational solutions.

The rest of the paper is structured as follows:  In section 2 we find the anomaly equations for general $q$ and review Lohitsiri and Tong's solution for the rational values of the standard model at $q=2$.  In section 3 we  consider the case of $q=3$ and rewrite the anomaly equations in Weierstrass form.  We discuss the torsion subgroups of the Mordell-Weil groups for the various values of $N$.   Using the Mordell-Weil theorem and Mazur's theorem we argue that the elliptic curves have infinitely many rational solutions for  $N\ne 3,9$ by showing that each curve has a set of rational solutions that cannot all fit into the torsion subgroup.  Thus, the rank of each curve is greater than zero.  For $N=3$  we find a set of  rational points that form a $\Z_9$ torsion subgroup and map these points to two inequivalent sets of hypercharges.  We then show that the nontrivial hypercharges are consistent with having a $\Z_2\times \Z_3$ center.  We do the same for $N=9$ where we find a set of  rational solutions that form a $\Z_{12}$ torsion subgroup.  Here the rational points split into three sets of hypercharges, two of which are nontrivial.  One set of hypercharges  can have a $\Z_2\times \Z_9$ center while the other can have a $\Z_2\times \Z_3$ center.   In section 4 we rewrite  the $N=3$ and $N=9$  curves in Weierstrass minimal form and find the curves in the Cremona tables, where the rank is explicitly shown to be zero in both cases. Hence, the only rational points are those that make up the torsion subgroups.   In section 5 we discuss the $L$-function for an elliptic curve and sketch how this is used by Cremona to determine the ranks for $N=3$ and $N=9$ by applying Kolyvagin's theorem, which proves part of the Birch Swinnerton-Dyer conjecture.  In the final section we discuss some further issues.

Arithmetic methods related to rational solutions have appeared recently in several other interesting physical applications. Rational sections have played an important part in the study of $F$-theory compactifications on elliptically fibered Calabi-Yau manifolds\cite{Aspinwall:1998xj}  (see \cite{Weigand:2018rez,Cvetic:2018bni} for reviews and an extensive list of references. For recent progress see \cite{Hajouji:2019vxs}.)  For example, a nonzero rank for the Mordell-Weil group of the fibration  leads to extra $U(1)$ gauge groups \cite{Morrison:2012ei,Cvetic:2013nia,Cvetic:2013qsa,Mayrhofer:2014opa,Kuntzler:2014ila,Esole:2014dea,Krippendorf:2015kta,Oehlmann:2016wsb}.  Rational sections in rational elliptic surfaces have also been studied in the Seiberg-Witten differentials and mass deformations of rank 1 $N=2$ supersymmetric quantum field theories\cite{Seiberg:1994aj, Minahan:1996fg, Minahan:1996cj, Noguchi:1999xq, Fukae:1999zs, Argyres:2015gha, Caorsi:2018ahl}. More relevant to the present work, rational solutions were also recently investigated for the anomaly equations of models with one or more $U(1)$ gauge groups or extra $U(1)$'s added to the Standard Model \cite{Batra:2005rh,Nakayama:2011dj,Cordova:2018cvg,Allanach:2018vjg,Rathsman:2019wyk,Costa:2019zzy}.

\section{The anomaly equations}

In this section we consider a generalization 
 to the standard model where the left-handed chiral multiplets are in  the {$\bf q$} representation of $SU(2)$ and the color group is assumed to be $SU(N)$.  In order to avoid $SU(N)^3$ anomalies there must be $q$  sets of anti-quarks in the anti-fundamental representation of $SU(N)$, each with an independent $U(1)$ hypercharge assignment $W_i$.  The hypercharges for the right-handed quarks, left-handed leptons and right-handed lepton are given by $Y_1$, $Y_4$ and $Y_5$ respectively.  Hence, we have the following representations:
\be
({\bf N},{\bf q})_{Y_1}, \ ({\bf \bar N},{\bf 1})_{W_1}\dots  ({\bf \bar N},{\bf 1})_{W_q}, \ ({\bf 1},{\bf q})_{Y_4}, \ ({\bf 1},{\bf 1})_{Y_5}\,.
\ee
In order to cancel the $SU(N)^2\times U(1)$, $SU(2)^2\times U(1)$ and $U(1)^3$ anomalies the charges must satisfy the equations
\be
qY_1+\sum_{i=1}^q W_i&=&0\nn\\
N Y_1+ Y_4&=&0\nn\\
Nq Y_1^3+N\sum_{i=1}^q W_i^3+q Y_4^3+Y_5^3&=&0\,.
\ee
Using the first two equations we can reduce the third equation to
\be\label{eq0}
(N^3-N)\left(\sum_{i=1}^q W_i\right)^3+Nq^2\sum_{i=1}^q W_i^3+q^2Y_5^3=0\,.
\ee
To ensure that there are no mixed $U(1)$ gravitational anomalies we also impose
\be\label{mixedgrav}
Nq Y_1+N\sum_{i=1}^q W_i+q Y_4+Y_5&=&0\,.
\ee
Hence, we can eliminate $Y_5$ and leave a cubic equation for the $W_i$,
\be\label{eq1}
-\left((q^2-1)N^2+1\right)\left(\sum_{i=1}^q W_i\right)^3+q^2\sum_{i=1}^q W_i^3&=&0\,.
\ee
Hence, assuming that not all hypercharges are zero, we are left with a cubic hypersurface in $\CP^{q-1}$.
If $q=2$ then the solutions are the isolated points on $\CP^1$.  To find them we rewrite (\ref{eq1}) as  
\be
-3(W_1+W_2)\left((N^2-1)(W_1^2+W_2^2)+2(N^2+1)W_1W_2\right)=0\,,
\ee
which has the solutions $(W_1,W_2)=(1,-1),\  (N+1,1-N), \ (1-N,N+1)$.  The last two are equivalent and correspond to the charge assignments for the standard model when $N=3$.  The first is an almost trivial solution but will continue to appear throughout this discussion.

Somewhat remarkably, for $q=2$ the mixed $U(1)$ gravitational anomaly forces the hypercharge assignments to be rational.  Rational charges are necessary if the $U(1)$ gauge theory is compact.  Even more remarkably, if  one assumes a compact $U(1)$ then the mixed anomalies automatically cancel \cite{Lohitsiri:2019fuu}!.  The argument is wonderfully simple.  If we substitute $W_1-W_2=U-V$ and $W_1+W_2=(U+V)/N$ then (\ref{eq0}) becomes
\be
4U^3+4V^3+4Y_5^3=0
\ee
which was proven by Euler to have no rational solutions, aside from the three trivial ones $(1,-1,0)$, $(0,1,-1)$ and $(-1,0,1)$, corresponding to the three solutions given above.  Notice that this argument does not rely on the value of $N$.

If we consider $q>2$, then clearly cancellation of the mixed anomalies is not enough to guarantee rational charges since the solution in (\ref{eq1}) is a surface with dimension $q-2$.  However, we can ask what happens if we require that the mixed anomalies cancel {\it and} the $U(1)$ gauge group is compact, in other words the charge assignments are rational.  We are not yet able to answer this question for general $q$, but for $q=3$ the results are quite interesting and make use of recent results in the study of elliptic curves.

\section{Elliptic equations}
\label{Elleq}

In this section we analyze the anomaly cancellation equation (\ref{eq1}) for $q=3$.  In this case we have the elliptic curve in $\CP^2$, given by
\be\label{elleq3}
-(8N^2+1)(W_1+W_2+W_3)^3+9(W_1^3+W_2^3+W_3^3)=0\,.
\ee
$\CP^2$ is the union of the affine plane and $\CP^1$, $\CP^2\simeq \mathbb{C}^2\cup \CP^1$.   One can see that the curve has a rational point on the $\CP^1$ where, say, $W_3=0$ and $W_1=-W_2$.  We call this point $\OO$.  There are obviously two other rational points where $W_1$ or $W_2$ are set to zero.  These rational  points have a clear $\Z_3$ symmetry where the zero is rotated into the different $W_i$.

At this point it is convenient to define a new set of variables,
\be
X&=&-W_3-\frac{8N^2+1}{8(N^2-1)}(W_1+W_2)\nn\\
Y&=&\frac{3}{2}(W_1-W_2)\nn\\
Z&=&-\frac{3}{8(N^2-1)}(W_1+W_2)\,.
\ee
With this substitution (\ref{elleq3}) becomes
\be\label{curveW}
ZY^2-X^3+3(8N^2+1)XZ^2-(16N^4+40N^2-2)Z^3=0
\ee
which can be put into Weierstrass form by setting $Z=1$,
\be\label{curve}
Y^2=X^3-3(8N^2+1)X+16N^4+40N^2-2\,.
\ee
The rational point $\OO$ corresponds to $Z=X=0$ and is included with the other rational points.
For a curve in the form $Y^2=X^3+BX+C$, the discriminant is $\Delta=-16(4B^3 + 27C^2)$.  Hence the discriminant of (\ref{curveW}) is 
\be\label{disc}
\Delta=-2^{12}\cdot 3^3 N^2(N^2-1)^3\,
\ee
which is nonzero if $N>1$, thus the elliptic curve in (\ref{curve}) is nonsingular.

We now invoke some useful facts about the rational points \footnote{For a nice textbook on this subject see \cite{Silverman_Tate}.  For reviews at a more advanced level see \cite{Silverman}, \cite{Milne}.}.  First the rational points are elements of an Abelian group called the Mordell-Weil group $E(\Q)$, which via the Mordell-Weil Theorem is a finitely generated  abelian group.   Hence, $E(\Q)$ has the form $E(\Q)=\Z^r\oplus T$ where $T$ is a finite subgroup called the torsion group and $r$ is  the rank.  Since $T$ is finite abelian it must consist of cyclic groups of finite order.  A theorem of Nagell and Lutz also says that  when the elliptic curve is written in Weierstrass form with $B$ and $C$ integer and $(X,Y)\in T$, then
$X$, $Y$ $\in\, \Z$ and either $Y=0$ or $Y^2|\Delta_0$, where $\Delta_0=-\Delta/16$.

The Mordell-Weil group is constructed as follows.  Take any two rational points $P$ and $Q$ and draw a line through them (assuming the points are distinct).  The line  either intersects the curve at a third point $R$, is tangent to either $P$ or $Q$, or is vertical and intersects the point $\OO$.  Since $P$ and $Q$ are assumed rational, then $R$ must  also be rational.  Let $\bar R$ be the point generated by reflecting $R$ about the $X$ axis, which is also a rational point on the curve because (\ref{curve}) is even in $Y$.  If we define the group element for a given rational point as $E(P)$ then the group multiplication is given by
\be\label{groupop}
E(P)\cdot E(Q)=E(\bar R)\,.
\ee
This is obviously Abelian.  It is also clear that $E(\OO)$ is the identity element and that $E(\bar R)=E(R)^{-1}$.  To carry out the group multiplication $E(P)\cdot E(P)$ choose the tangent to the curve at $P$ as the line.  This line will intersect the curve at a point $R$ and the result of the group multiplication is $E(\bar R)$.  It is clear from this that the only points of order two are $\OO$ and  rational points on the $X$-axis.

From this construction we  see that (\ref{curve}) has at least a $\Z_3$ torsion subgroup corresponding to the rational points described below (\ref{elleq3}), which in terms of the coordinates in (\ref{curveW}) are the points  $\OO$ and $(X,Y)=(3,\pm4(N^2-1))$.
We now can use a theorem of Mazur on the allowed torsion subgroups for an elliptic curve\cite{Mazur}.  Mazur proved that the only possible torsion subgroups are $\Z_n$ with $1\le n\le10$ and $n=12$, and $\Z_2\oplus \Z_{2n}$ with $1\le n\le 4$.  In fact the latter subgroups are only possible if $\Delta>0$.  Since  the curve in (\ref{curve}) must contain a $\Z_3$ subgroup and its discriminant is negative, it then follows that the only possible torsion subgroups for these curves are $\Z_3$, $\Z_6$, $\Z_9$ or $\Z_{12}$.  This is enough information to show that except for $N=3$ and $N=9$, the rank satisfies $r>0$.  

To show this, we first note that the possible orders of the torsion subgroup are $3$, $6$, $9$ and $12$.   Hence, if we find more than 12 distinct integer points, including $\OO$, then at least one of the points cannot be entirely in the torsion subgroup, and hence $r>0$.  To this end it is straightforward to show that the 14 integer points 
\be\label{14points}
(X,Y)&=&\big(3,\pm4(N^2-1)\big),\ \big(4N-1,\pm4N(N-1)\big),\ \big(8N+11,\pm4(N+1)(N+9)\big),\nn\\
&&\big(N^2+2,\pm N(N^2-1)\big),\ \big(4N^2-1,\pm8N(N^2-1)\big),\ \big(-4N-1,\pm4N(N+1)\big),\nn\\
&&\big(-8N+11,\pm4(N-1)(N-9)\big)
\ee
are all solutions to (\ref{curve}).  Including $\OO$ this gives at least 15 integer solutions.  
The only exception is if some of the points are duplicated.  Since the first 5 sets of points all have $X>0$ and the last two have $X<0$ for $N\ge 2$ we need to only check within these subsets of points for duplicates.  Since the last two equations have an $X$ coordinate linear in $N$, they can cross only at one value, which is $N=3$.  Likewise, it is not hard to show that the duplicates for $X>0$ are for $4N-1=N^2+2$ and $8N+11=4N^2-1$ when $N=3$, and $8N+11=N^2+2$ when $N=9$.  The only other duplication occurs when $N=9$ where the last set has $Y=0$, hence there is only one solution here, not two.  With these duplications, we see that for $N=3$ this reduces the points in (\ref{14points}) plus $\OO$ to  9 distinct points, while for $N=9$ the reduced number is 12.  Both are consistent with the order of a torsion subgroup.
\footnote{A simpler way to argue that  $r>0$ for all $N$ except $N=3,6,9$ is to use the Nagell-Lutz theorem.  We observe that $Y^2\nmid\Delta_0$ if $Y=\pm4(N-1)(N-9)$ or $Y=\pm4(N+1)(N+9)$, unless $N=3,6,9$.  Hence such points are not contained within the torsion subgroup and must have infinite order. Furthermore, by applying the duplication formula to the point $(59, 420)$, we can get a non-integer rational point $(299/25, 14532/125)$ and therefore conclude that $r > 0$ for  $N=6$.}

A  useful fact about the torsion subgroup for a general elliptic curve is that there exists an injective homomorphism from the  torsion subgroup into the group generated by the solutions of the elliptic curve over the finite field $\F_p$, where $p$ is any prime over which the curve has a ``good reduction".  This last point means that the curve is not singular over $p$ which is the case if its discriminant  $\Delta$ is not divisible by $p$.  Thus,  (\ref{curveW}) has a good reduction for any $p$ except $p=2,3$ and those primes that divide $N-1$, $N$, or $N+1$.  
Since the homomorphism is injective, then ${\rm ord}(T)|\# E(\F_p)$, where ${\rm ord}(T)$ is the order of $T$ and $\# E(\F_p)$ is the number of solutions over $\F_p$.   We can then use this to find $T$,   usually by just checking for a few primes.  For example, for $N=4$ (\ref{curve}) has good reduction for $p\ge 7$ and one finds that $\#E(\F_7)=9$, $\#E(\F_{11})=18$ and $\#E(\F_{13})=12$.  Since ${\rm ord}(T)$ must divide all three numbers we see that it is at most $3$.  Since we know that there is at least a $\Z_3$ torsion subgroup, the torsion subgroup must be $\Z_3$.  

However, for $N=3$ (\ref{curve}) has good reduction for all primes aside from $2$ and $3$.  Here one finds for the first several hundred primes, starting at $p=5$, that $9|\#E(\F_p)$, strongly suggesting that $T\simeq \Z_9$.  In fact, using the group operations in (\ref{groupop}) it is a straightforward exercise to show that the eight distinct points in (\ref{14points}) and $\OO$ make up the elements of $\Z_9$.  Six of the points, $(11,\pm 24)$, $(-13,\pm45)$ and $(35,\pm192)$ are points of order 9 and map to the $W_i$ values $(W_1,W_2,W_3)=(1,-5,7)$, plus permutations. The other hypercharges are then $(Y_1,Y_4,Y_5)=(-1,3,-9)$. The other three rational points make up the $\Z_3$ subgroup and map to the almost trivial solution $(W_1,W_2,W_3)=(1,-1,0)$ up to permutations.  Hence, up to relabeling, there is one nontrivial rational  solution, assuming that $r=0$.

If  $N=9$ then  (\ref{curve}) has  good reduction except for $p=2,3,5$.  Starting at $p=7$ one finds that $12|\#E(\F_p)$  for the next one hundred primes, consistent with $T\simeq \Z_{12}$.  Using the group operation in (\ref{groupop}) it is again straightforward to show that the 11 distinct points in (\ref{14points}) and $\OO$ make up the elements of $\Z_{12}$.     These include an order 2 point at $(-61,0)$, an order 4 point at $(35,288)$, an order 6 point at $(83,720)$ and an order 12 point at $(323,5760)$.  The four points of order 12 and the two points of order 4 correspond to the hypercharge assignments $(W_1,W_2,W_3)=(1,-17,19)$ plus permutations, while the elements of order 6 and 2 map to $(W_1,W_2,W_3)=(5,5,-13)$.  The other hypercharge assignments are $(Y_1,Y_4,Y_5)=\pm(-1,9,-27)$.  Hence, we find two nontrivial solutions, up to relabeling, assuming that $r=0$.

In both the $N=3$ and $N=9$ case the nontrivial hypercharge assignments are consistent with a center symmetry which acts nontrivially on the $SU(N)$ and $SU(2)$ representations.   Since the $SU(2)$ triplet is even under the $\Z_2$ center the multiplets are invariant under it.  In the case of the $\Z_N$ center of $SU(N)$ it needs to be accompanied by a $\Z_N$ transformation in the $U(1)$.  One can check that for the nontrivial $N=3$ solution and the first nontrivial $N=9$ solution that
\be\label{centercond}
\frac{c_R}{N}+\frac{W_i}{N}\in \Z
\ee
where $c_R$ is the $N$-ality of the $SU(N)$ representation.  Hence the center symmetry $\Gamma$ is $\Gamma=\Z_2\times \Z_N$ in these cases.  For the second $N=9$ solution one finds that (\ref{centercond}) is not satisfied but it is if the lefthand side is multiplied by 3.  Hence this supports a $\Gamma=\Z_2\times \Z_3$ center symmetry.  Because of these center symmetries the gauge groups can be reduced to $SU(N)\times SU(2)\times U(1)/\Gamma'$ where $\Gamma'$ is $\Gamma$ or one of its subgroups.  If $\Gamma'$ contains the $\Z_2$, then since this symmetry is diagonal on the $SU(2)$ it reduces this part of the gauge group to $SO(3)$.
  
There are many routines available to compute the torsion subgroups for general $N$, including on SageMath.  These take advantage of the Nagell-Lutz theorem, Mazur's theorem and the property that $\rm{ord}(T)|\# E(\F_p)$ to rapidly determine $T$.   They can quickly run through the first $10^6$ values of $N$ and show that for $N\ne3,9$ the torsion subgroup is $T\simeq \Z_3$, unless $N=(m+1)(m^2+2m-2)/2$, $m\ge3$, in which case $T\simeq \Z_6$.  In this latter case the curve has a rational point at  $(-m^4-4m^3-4m^2+3,0)$. Since this is on the $X$-axis it corresponds to a group element of order 2.  Combining with the $\Z_3$ subgroup this gives  $\Z_6\simeq \Z_2\oplus \Z_3$.

\section{Results from the Cremona tables}

It is much more  challenging to find the rank. Fortunately, Cremona maintains a large database of elliptic curves that explicitly provides the torsion subgroup and the rank  for a large class of elliptic curves \cite{cremona} \footnote{A searchable frontend for the database is found at \href{http://www.lmfdb.org/EllipticCurve/Q/}{{\tt http://www.lmfdb.org/EllipticCurve/Q/}}}.   The curves in (\ref{curve}) for $N=3$ and $N=9$ are in this database, where the torsion subgroups are $\Z_9$ and $\Z_{12}$ respectively, and the ranks for both are found to be zero.  Hence for these values of $N$ the only rational solutions are those listed in the previous section.  In this section we provide further details about these curves.  In the following section we sketch how the rank is determined \cite{cremonabook} using Kolyvagin's theorem \cite{Kolyvagin} for part of the Birch Swinnerton-Dyer conjecture \cite{BS-D}.

Elliptic curves have associated with them an integer invariant known as the conductor, $N_c$ \footnote{In the mathematics literature the conductor is usually written as $N$, but we use $N_c$ to distinguish it from the number of colors.}.  The elliptic curves in \cite{cremona} are ordered starting at the smallest possible value of $N_c$.  To find $N_c$, one first needs to put (\ref{curve}) into Weierstrass minimal form.   
To do this one shifts and rescales the coordinates, transforming the curve to
 \be\label{curvemin}
 y^2+a_1 xy+a_3 y=x^3+a_2x^2+a_4 x+a_6\,,\qquad a_i\in \Z
 \ee
 such that the discriminant is minimized.  Such transformations must give back an integral discriminant, and  furthermore must reduce the value of the discriminant by an integer to the twelfth power.  The fact that (\ref{disc}) has an explicit factor of $2^{12}$   suggests that (\ref{curve}) can be reduced.  Indeed for general odd $N$ the minimal curve is
 \be\label{curveodd}
 y^2+xy+y=x^3-x^2-\frac{3N^2+1}{2}x+\frac{N^4+4N^2-1}{4}\,,
 \ee
 while for even $N$ one has
 \be\label{curveeven}
 y^2+xy=x^3-x^2-\frac{3N^2}{2}x+\frac{N^4+4N^2}{4}\,.
 \ee
 The discriminant for curves in the form (\ref{curvemin}) is given in terms of a set of standardized coefficients for the curve.  These are
 \be\label{ECcoeffs}
&& b_2=a_1^2+4a_2\,,\qquad b_4=a_1a_3+2a_4\,,\qquad b_6=a_3^2+4a_6\nn\\
&&b_8=a_1^2a_6-a_1a_3a_4+4a_2a_6+a_2a_3^2-a_4^2\nn\\
&&c_4=b_2^2-24b_4\,,\qquad c_6=-b_2^3+36b_2b_4-216b_6\,.
\ee
In terms of these quantities the discriminant is
\be\label{discmin}
\Delta=\frac{c_4^3-c_6^2}{1728}\,.
\ee
For the curves in (\ref{curveodd}) and (\ref{curveeven}) we find
\be\label{ECoeffsN}
&&b_2=-3\,,\qquad b_4=-3N^2\,,\qquad b_6=N^2(N^2+4)\,,\qquad b_8=-3N^2(N^2+1)\,,\nn\\
&&c_4=9(8N^2+1)\,,\qquad c_6=-27(8N^4+20N^2-1)\,,\nn\\
&&\Delta=-3^3N^2(N^2-1)^3\,.
\ee
   While the discriminant in (\ref{ECoeffsN}) is smaller than  (\ref{disc}), it is still even and so contains the same prime factors.  Hence, it does not change whether or not the curve has a good or bad reduction for any $p$.  If $N=27m\pm 1$ then (\ref{curvemin}) and (\ref{curveodd}) can be  reduced even further
to
\be\label{oddm}
Y^2+XY+Y=X^3-\frac{27m^2\pm2m+1}{2}X+\frac{(27m^2\pm2m+1)(27m^2\pm2m-1)}{4}\,
\ee
for $m$ odd and
\be\label{evenm}
Y^2+XY=X^3-\frac{27m^2\pm2m}{2}X+\frac{(27m^2\pm2m)^2}{4}\,
\ee
for $m$ even.  The discriminant of these new curves is $\Delta=-m^3(27m\pm1)^2(27m\pm2)^3$.  If $m\ne 0\mod 3$ then $3$ does not divide the discriminant and so $p=3$ now has  good reduction.

To get the conductor we need to further distinguish between different types of bad reduction over $\F_p$.  A singular elliptic curve can have two types of singularities, a node or a cusp.  A node occurs where the curve crosses itself with distinct tangents.  A cusp occurs if the tangents are equal.  If we consider the curve over $\F_p$, the singularity is a node if $p\nmid c_4$.  In this case the reduction is called multiplicative.  Since $c_4$ is always odd, we see that there is a node over $\F_2$.  For $p>3$ that divides $N-1$, $N$ or $N+1$ it is also easy to show that $p\nmid c_4$, hence these also have multiplicative reduction.  If $p=3$ then we see that $p|c_4$ in (\ref{ECoeffsN}), in which case the singularity is a cusp.  Such a reduction is called additive.

For every prime $p$ we then assign a nonnegative integer $f_p$ where
\begin{itemize}
\item $f_p=0$ if $p\nmid \Delta$;  good reduction
\item $f_p=1$ if $p|\Delta$, $p\nmid c_4$; multiplicative  reduction
\item $f_p=2 +\delta_p$ if $p|\Delta$, $p| c_4$, $p>3$; additive  reduction
\end{itemize}
For the last case $\delta_p=0$ if $p>3$.
The conductor is then defined as 
\be
N_c=\prod_p p^{f_p}
\ee
 To find $\delta_3$ one can use the Tate algorithm (see \cite{SilvermanII}, IV.9 for an explanation).  Here we  state without further explanation that $f_3=3$ for the $N=3$ curve while  $f_3=2$ for $N=9$.  Hence the conductors for $N=3$ and $N=9$ are $N_c=54$ and $N_c=90$ respectively.  These are quite low and the curves are  readily found in the Cremona tables.  In particular, the $N=3$ curve in minimal form is the elliptic curve {\bf 54b3}, while the $N=9$ curve is {\bf 90c3} \footnote{Note that the cubic in (\ref{cubic}) is equivalent to {\bf 27a1} in the tables.}.  Both curves have rank 0.  It is also interesting that these curves have the smallest possible $N_c$ for curves with a $\Z_9$ and $\Z_{12}$ torsion respectively.
  
  The curves for all other values of $N$ which have $N_c\le 400,\!000$ are also contained in the tables.  This includes every curve with $N<34$, all being confirmed to have $r\ge1$ if $N\ne3,9$.  The smallest $N$ where $r=2$ is at $N=18$.  Using SageMath to go above $N=34$ we find the first rank 3 curve at $N=93$.

\section{The $L$-function and the Birch Swinnerton-Dyer conjecture}

In this section we briefly summarize a powerful method used to determine the ranks in \cite{cremona}.     Further details can be found in \cite{cremonabook}.

We first define the $L$-function for an elliptic curve $E$.  Consider the curve  in minimal form over the finite field $\F_p$ which has good reduction and define $a(p)\equiv 1+p-A(p)$, where $A(p)$ is the number of solutions, including the point at infinity.  For example, for the $N=3$ curve in (\ref{curveodd}), $A(5)=9$ and so $a(5)=-3$.    On average, for any give $X$ coordinate mod $p$, we expect the curve to cross one value of $Y$ mod $p$.  Hence, including $\OO$ we expect an average of $p+1$ solutions.  $a(p)$  essentially measures the deviation away from this average and by Hasse's theorem is bounded by $|a(p)|\le2\sqrt{p}$.

We then define the $L$-function for an elliptic curve $E$ as
\be\label{Lf}
L(E,s)=\prod_{p}\left(1-a(p)p^{-s}+\chi(p)p^{1-2s}\right)^{-1}\,.
\ee
where $\chi(p)=1$ if $p\nmid N_c$ and $\chi(p)=0$ otherwise.  In this last case, if the curve has additive reduction over $p$ then $a(p)=0$, while if the curve has multiplicative reduction over $p$ then $a(p)=-\eps_p$, where $\eps_p=\pm1$.  If the multiplicative reduction is split,  which means that the slopes of the tangents are in $\F_p$, then  $\eps_p=-1$, while $\eps_p=+1$ for non-split reduction.  For $p>3$ the reduction is split if $-c_6$ is a square in $\F_p$.  For the curves in (\ref{curveodd}) and (\ref{curveeven}), if $p|(N\pm1)$ the reduction is split because $-c_6=27^2 \mod p$.  For $p|N$ the reduction is split if $-27 \mod p$ is a square.  In the case of $p=2$ it turns out that there is split reduction for $N$ odd and non-split for $N$ even.  

Expanding out the product in (\ref{Lf}) we then express $L(E,s)$ as a Dirichlet series,
\be\label{Ds}
L(E,s)=\sum_{n=1}^\infty a(n) n^{-s}\,,
\ee
which converges if $\Re(s)>3/2$ because of the Hasse bound. 
We then write $L(E,s)$ as the Mellin transform of the function $f(E,z)$
\be\label{Mellin}
L(E,s)=\frac{(2\pi)^s}{\Gamma(s)}\int_0^{i\,\infty}(-iz)^sf(E, z)\frac{dz}{z}\,,
\ee
where $f(E,z)$ is given by
\be\label{fz}
f(E,z)=\sum_{n=1}^\infty a(n)e^{2\pi i n z}\,.
\ee
With the transform in (\ref{Mellin}) we can analytically continue to $s=1$.  The Birch Swinnerton-Dyer conjecture then states that $L(E,s)$ has a Taylor series expansion about $s=1$ of the form
\be\label{BS-Dconj}
L(E,s)=C (s-1)^r+\dots\,,
\ee
where $r$ is the rank of the curve and $C$ is a nonzero number composed of other invariants of the curve.  Kolyvagin proved the conjecture is true for $r=0$ and $r=1$ \cite{Kolyvagin}, following work of Gross and Zagier who showed that if $r=1$ then the elliptic curve had infinitely many rational points \cite{GZ}.  Hence, if one can show that $L(E,1)\ne0$ then the curve must have rank 0.

The coefficients $a(n)$ in (\ref{Ds}) and (\ref{fz}) have the following composition properties:
\begin{itemize}
\item $a(nm)=a(n)\,a(m)$ if gcd$(n,m)=1$
\item $a(p^{q+1})=a(p)\,a(p^q)-p\, a(p^{q-1})$ if $p\nmid N_c$
\item $a(p^{q+1})=a(p)\,a(p^q)$ if $p|N_c$,
\end{itemize}
These conditions are required for $f(E, z)$ to be a newform, a cusp form of weight 2 over the congruent subgroup $\Gamma_0(N_c)$ of $SL(2,\Z)$ which is not a cusp form of $\Gamma_0(M)$ where $M|N_c$.  Note that the elements of $\Gamma_0(N_c)$ are $\left(\begin{array}{cc} a&b\\ c&d\end{array}\right)$ where $ad-bc=1$ and $c=0 \mod N_c$. Every isogeny class of elliptic curves\footnote{An isogeny is a rational map  that is also a homomorphism of the point multiplication group.} has such a  newform associated with it \cite{BCDT}.

Crucially, $f(E, z)$ is an eigenfunction of the Fricke involution, the  Atkin-Lehner transformation given by
\be
W_{N_c}f(E, z)\equiv \frac{1}{N_cz^2}f(E, -1/N_cz)=\eps\,  f(E, z)\,.
\ee
Since $W_{N_c}$ is an involution the eigenvalue must satisfy $\eps=\pm1$ .  Using this one can argue that 
\be\label{transint}
L(E,s)&= &\frac{(2\pi)^sN_c^{-s/2}}{\Gamma(s)}\int_0^{\infty}(y)^sf(E, iy/\sqrt{N_c})\frac{dy}{y}\nn\\
&=& \frac{(2\pi)^sN_c^{-s/2}}{\Gamma(s)}\int_1^{\infty}\left(y^s-\eps y^{2-s}\right)f(E, iy/\sqrt{N_c})\frac{dy}{y}\,.
\ee
If $\eps=+1$ then the integrand in (\ref{transint}) is odd under $s\to 2-s$ and hence $r$ in (\ref{BS-Dconj}) is odd, while if $\eps=-1$ then $r$ is even.  It turns out that for the $N=3$ and $N=9$ curves  $\eps=-1$.  In these cases we can then write $L(E,1)$  as
\be
L(E,1)=2\sum_{n=1}^{\infty} \frac{a(n)}{n}\,e^{-2\pi n/\sqrt{N_c}}\,.
\ee
If $N_c$ is not too large then the sum converges rapidly and one can get very accurate results with only a handful of $a(n)$ values.  For $N=3$ the result is $L(E,1)=1.0305218\dots$, while for $N=9$ it is $L(E,1)=1.3375999\dots$.  Hence the rank for both is zero.
\section{Discussion}

The models we have discussed for $q>2$ are not phenomenologically viable, nor do they seem to fit  into  larger grand unified models in a nice way.  But their supersymmetric versions might be found in  string theory compactifications, keeping them out of the Swampland.  
 String compactifications admitting gauge theories with matter multiplets in higher representations have been constructed in the past, for example using higher level Kac-Moody algebras for  heterotic compactifications that are free on the world-sheet \cite{Lewellen:1989qe,Dienes:1996yh,Dienes:1996wx}.  It is also possible to get adjoint matter multiplets in IIA string theory,  and higher representations seem to be possible in $F$-theory but this has not been seriously explored   \cite{Halverson:2018xge}.  It would be interesting to  carry out such a search for these types of models, perhaps using the ideas in \cite{Halverson:2017ffz,Taylor:2015xtz,Taylor:2017yqr} .
 
It also would be interesting to see if any new rational solutions are allowed if one drops the mixed gravitational anomaly condition in (\ref{mixedgrav}).  That the mixed gravitational anomalies give rational charges for $q=2$ might not be accidental.  If one considers a noncompact Abelian hypercharge group and two matter fields with  irrational hypercharge ratios, then there must be a nontrivial global  $U(1)$ symmetry in the theory \cite{Banks:2010zn}.  Global symmetries cannot  exist in consistent theories of quantum gravity \cite{Banks:1988yz,Banks:2010zn}, so it is satisfying that once the mixed gravitational anomalies are included the ratios are forced to be rational.  This will not happen if $q>2$, but the converse might be true.  That is, all allowed rational solutions are consistent with cancellation of the mixed gravitational anomalies.  This is not guaranteed to work since there are known examples with a compact $U(1)$ gauge group which are free of gauge anomalies but have mixed gravitational anomalies \cite{Lohitsiri:2019fuu}.

\section*{Acknowledgements}

This research  is supported in part by Vetenskapsr{\aa}det under grant \#2016-03503 and by the Knut and Alice Wallenberg Foundation under grant Dnr KAW 2015.0083. YL  thanks Nordita for kind hospitality during the course of this work. JAM also thanks the Center for Theoretical Physics at MIT and Nordita for kind hospitality.

\bibliographystyle{JHEP}
\bibliography{anom_ell}  
 
\end{document}